\documentclass[twosides]{article}
\usepackage{amsmath}
\usepackage{amssymb}
\usepackage{psfrag}
\usepackage[dvips]{graphicx}
\usepackage[dvips]{color}
\usepackage{latexsym,amsfonts}
\usepackage{multicol}

\begin{document} 

\title{Kinematics and Quantum Field Theory of the Neutrino Oscillations Observed in the Time-modulated Orbital Electron Capture Decay in an Ion Storage Ring}
 
\author{
Manfried Faber\thanks{\mbox{\it E-mail}: faber at kph.tuwien.ac.at}\\[5mm]
Atominstitut der \"Osterreichischen
    Universit\"aten, Technische Universit\"at Wien,\\
    Wiedner Hauptstra\ss e 8-10, A-1040 Wien, \"Osterreich}
 

\date{\today}

\maketitle

\begin{abstract}
According to the recent experimental data of a GSI--experiment, the rate of the number of daughter ions $^{140}{\rm Ce}^{58+}$, produced by a nuclear K--shell electron capture (EC) of the H--like ion ${^{140}}{\rm Pr}^{58+}$, is modulated in time with a period $T_d=(7.06\pm 8)\,$seconds. We explain this phenomenon by neutrino--flavour mixing and show that this can be understood within standard quantum field theory and derive a value for the squared mass difference $\Delta m^2_{21}=m^2_2-m^2_1=(0.763\pm 8)\,\cdot 10^{-4}\,\mathrm{eV}^2$. This proves that such processes provide a precise method to investigate neutrino--flavour mixing.
\end{abstract}

\begin{center}
\end{center}

\maketitle

\section{Introduction}
\begin{multicols}{2}
Recently, K--electron capture (EC) and $\beta^+$ decays of H--like $^{140}\mathrm{Pr}^{58+}$ ions were measured in the Experimental Storage Ring (ESR) at GSI in Darmstadt \cite{Litvinov:2007ck}. It is very exciting that in this electron capture process periodic modulations of the expected exponential decrease of the number of EC--decays per time was observed \cite{Litvinov:2008rk}. The rate $\mathrm dN^\mathrm{EC}_d(t)/\mathrm dt$ of the number of daughter ions $^{140}\mathrm{Ce}^{58+}$ produced in EC--decay $^{140}\mathrm{Pr}^{58+} \to ^{140}\mathrm{Ce}^{58+} + \nu$  shows a modulation in time with a period $T_d = (7.06\pm 8)$~seconds \cite{Litvinov:2008rk}. This value of $T_d$ is the result of a fit  with the function
\begin{eqnarray}\label{activity} 
\frac{\mathrm dN^\mathrm{EC}_d(t)}{\mathrm dt}=\lambda_\mathrm{EC}(t)\,
\mathrm e^{-\lambda t}N_m(0)
\end{eqnarray}
where $N_m(0)$ is the number of mother ions at the production time $t=0$, the decay constant $\lambda$ includes loss of particles and $\beta^+$ decay. The time dependent ``decay factor'' $\lambda_\mathrm{EC}(t)$, one cannot call it anymore decay constant, was fitted by
\begin{align}\label{Lambdat}
\lambda_\mathrm{EC}(t)=\bar\lambda_\mathrm{EC}[1+a\cdot \cos(\Omega t+\phi)]
\end{align}
with a result of the fit, $\Omega=2\pi/T_d=0.89(1)~\mathrm s^{-1}$ and $a=0.15(3)$.

It was shown in \cite{Ivanov:2007pp} that the K-electron capture to positron emission ratios of H--like $^{140}\mathrm{Pr}^{58+}$ and the He--like $^{140}\mathrm{Pr}^{57+}$ ions can be described within the standard theory of weak interactions of heavy ions \cite{ST2} with an accuracy better than $3\,\%$.

In recent years neutrino oscillation predicted by Bruno Pontecorvo \cite{Pontecorvo:1957qd} have witness great progress both experimentally and theoretically \cite{Yao:2006px},\cite{Grimus:2003es}. During the last decades experimental evidence for neutrino oscillations came from solar neutrino experiments (Homestake, Kamiokande, SAGE, GALLEX-GNO, SuperKamiokande (SK) and SNO) long-baseline reactor neutrino experiment (KamLAND), atmospheric neutrino experiments (SK, MACRO, and Soudan-2) and long-baseline accelerator neutrino experiment (K2K)\cite{Yao:2006px}.

In this article we would like to support the idea that neutrino oscillations can also be detected in electron capture processes. We show that the observed time--modulation of the number of daughter nuclei in the EC--decay of the H--like $^{140}\mathrm{Pr}^{58+}$ ions can be explained by neutrino--flavour mixing \cite{Maki:1962aa}. Therefore, EC--decays provide a new method for studying neutrino mixing and mass differences of neutrino flavours.

In the following we will first discuss the kinematics of the process, then we will start the quantum theoretical calculations with the calculation of the transition matrix element and finally we will determine the decay rate by the methods of quantum field theory. These calculations should prove that we have picked up the right kinematics. In the main part of the text the most important steps of the calculations are shown.\footnote{The detailed quantum theoretical calculations are shifted to the appendix.} After these investigations we discuss the relations for the amplitude of the oscillations as well as some features of the results.

\section{Kinematics}
In the K-capture the mother ion $^{140}\mathrm{Pr}^{58+}$ makes a transition to the daughter nucleus $^{140}\mathrm{Ce}^{58+}$ with the mass $M_d=130319.252\,\mathrm{MeV}$ and the Q--value
\begin{align}\label{QValue}
Q=M_m-M_d= 3348\pm 6\,\mathrm{keV}.
\end{align}
In the rest frame of the mother ion $^{140}\mathrm{Pr}^{58+}$ we are allowed to take the non--relativistic expressions for the energy of the daughter nucleus. If a neutrino species of mass $m_j$ is emitted we expect due to energy and momentum conservation the relation
\begin{align}\label{EneCons}
\sqrt{k_j^2+m_j^2}+\frac{k_j^2}{2M_d}=Q,
\end{align}
where $k_j$ is the momentum of the neutrino in the mass eigenstate $m_j$. Inserting into this equation $j=1$ and $2$, we expect from the best--fit
\begin{align}\label{Dm2GF}
\Delta m^2_{21}=m_2^2-m_1^2=8.0^{+0.6}_{-0.5}\cdot 10^{-5}\,\mathrm{eV}^2
\end{align}
of the global analysis of the solar--neutrino and KamLAND experimental data \cite{Yao:2006px} (see also \cite{Fogli:2006aa}) momenta of $k_1\approx k_2\approx Q=3.388\cdot 10^6\,\mathrm{eV}$ and a tiny momentum difference of
\begin{align}\label{Delk}
k_1-k_2=1.12\cdot 10^{-11}\,\mathrm{eV},
\end{align}
which can be estimated to be about $k_1-k_2\approx\Delta m_{12}^2/(2Q)$. For the further kinematic calculation we use two neutrino species only. The quantum theoretical part we will do with a general set of neutrino mass eigenstates. In the standard procedure for the determination of the decay rate one integrates over the momenta $\vec q$ of the daughter nuclei with mass $M_d$ or the momenta $\vec k_j$ of the emitted neutrino with mass $m_j$. After some approximations this could lead to a guess of the oscillation period $T$ caused by the mixing of the neutrino mass eigenstates $m_1$ and $m_2$ of  $T_{12}=2\pi/(k_1-k_2)\approx 2\pi\cdot 2Q/\Delta m_{12}^2=0.3\,\mathrm{ms}$ which is far away from the experimental value, $T_d\approx 7\,\mathrm s$ and is disproved by the experiment.

One can get much closer to the experimental value of $T_d$ using an oscillation frequency $\Omega_{12}=\Delta m_{12}^2/(2M_d)$ which one can guess characterising the emitted neutrinos by their energy $E_\nu$ and using energy and momentum conservation for neutrinos of different mass and the same energy $E_\nu$. This idea leads to the energy and momentum conservation relation $E_\nu+(E_\nu^2-m_j^2)/(2M_d)=Q$ and to the prediction $T_{12}=13\,\mathrm s$ \cite{Litvinov:2008rk}. But this idea suffers from the problem how to integrate over the final neutrino states. In the standard procedure one has to integrate over the momenta of the final states, but the two neutrinos with the same energy and with different masses which should interfere have two phase spaces of different size. Which size does one have to use for the integration? Even if the momentum difference (\ref{Delk}) is tiny compared to the absolute values $k_1\approx k_2\approx Q$ of the momenta, this problem has to be solved.

This difficulty with the definition of the phase space shows that two involved mass--eigenstates, a lighter $m_j$ and a heavier $m_l$, need exactly the same momentum $\vec k$ in the creation process when they are emitted from the mother nucleus. At the same time the emitted neutrinos have to have those momenta $k_j$ and $k_l$ and energies which follow from the energy and momentum conservation relations (\ref{EneCons}) in the centre of mass system of the mother ion. In classical mechanics this looks like a contradiction. In quantum mechanics this is possible, since the invention of Schr\"odingers cats we are used to have in quantum mechanics objects in a superposition of different states.

\end{multicols}
\rule{64mm}{.1pt}
\begin{figure}[h]
\psfrag{k1}{$\vec k_j$}\psfrag{k2}{$\vec k_l$}\psfrag{k}{$\vec\kappa$}
\psfrag{f}{$\varphi$}\psfrag{A}{$A$}
\centering
\includegraphics[scale=0.75]{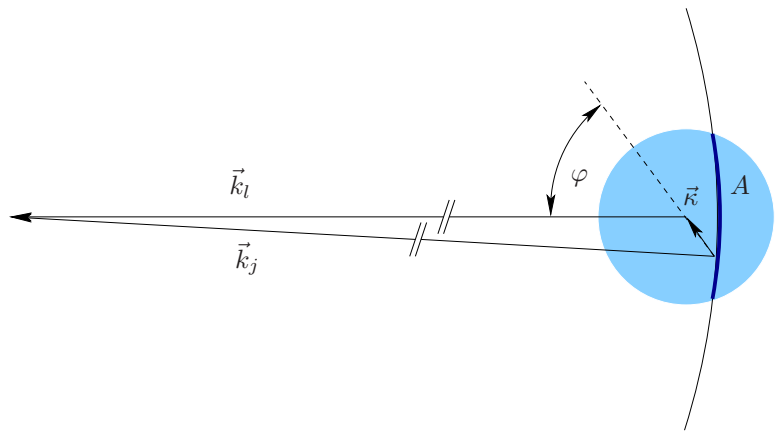}
\caption{Scheme of the momenta involved in the kinematics. The momentum uncertainty $\vec\kappa$ of the mother nucleus and the momentum $\vec k_l$ of the neutrino mass eigenstate with mass $m_l$ add up to the momentum $\vec k_j$ of the neutrino with mass $m_j$. The order of magnitude of the radius $k_j$ of the sphere is by many magnitudes larger than the magnitude of $-\kappa$. Therefore the sphere with the radius $\vec k_j$ is rather a plane. $A$ is the cross--section of this plane with the shaded region of allowed $\vec\kappa$ values. If $A$ shrinks to zero there are no oscillations.}
\label{kinem}
\end{figure}
\begin{multicols}{2}
\vspace*{-13mm}\hspace{75mm}\rule{64mm}{.1pt}
For the two neutrino mass eigenstates the momenta in the centre of mass system differ by a momentum $\vec\kappa=\vec k_j-\vec k_l$, see Fig.~\ref{kinem}. According to Eq.~(\ref{EneCons}) the lighter neutrino species has the larger momentum. Due to the experimental arrangement in the GSI experiment~\cite{Litvinov:2008rk} with ions in a storage ring the mother ions have some momentum uncertainty and we have to assume a wave--packet for the mother ions with a width $\sigma_\kappa$ for the distribution of the total momentum $-\vec\kappa$. Then neutrinos can be emitted from any component of this wave--packet moving against the centre of mass system $S_0$.

For simplicity we assume that the mass eigenstate $m_j$ is emitted in $S_0$ with momentum $\vec k=\vec k_j$ conserving energy and momentum (\ref{EneCons}). To get the same momentum for $m_l$ in the emission process we have to emit it from a component of the mother nucleus which moves with momentum $-\vec\kappa=-(\vec k_j-\vec k_l)$ backwards. In this system $S_{-\kappa}$ energy and momentum are conserved for the emission of $m_2$ (at the vertex of the Feynman diagram). 

Then, in the centre of mass system $S_0$ the energy conservation is violated for the emission of $m_l$, though $m_l$ is moving with momentum $\vec k_l$. It is the daughter nucleus which moves too fast, with momentum $-(\vec k_l+2\vec\kappa)$. The energy conservation is violated by a small amount $\Delta E_l$. The amount of this violation gives a period $T_{jl}$ of the oscillation of the number of decaying nuclei by $T_{jl}=2\pi/\Delta E_l$.

The value of the energy violation $\Delta E_l$ for the neutrino with mass $m_l$ as a function of the neutrino momentum $k$, the momentum $\kappa$ of the mother ion and the angle $\varphi$, see Fig.~1, between the two momenta is given by
\begin{align}\label{OffShellE}
&\Delta E_l(k,\kappa,\varphi)=\\
&=\sqrt{(\vec k-\vec\kappa)^2+m_l^2}+\frac{(\vec k+\vec\kappa)^2}{2M_d}
-Q-\frac{\vec\kappa^2}{2M_m}.\nonumber
\end{align}
With Eqs.~(\ref{EneCons}) and (\ref{OffShellE}) we can predict from the oscillation period $T_d$ in the GSI--experiment the difference of the squared masses of the corresponding neutrinos. Due to the very different scales involved in this process the calculations have to be done with an accuracy of at least 25 digits. The scales range from the mass $M_d$ in the order of $130\,\mathrm{GeV}$ to the energy violation, in the order of $6\cdot 10^{-16}\,\mathrm{eV}$, i.e. over $27$ orders of magnitude.

For simplicity we start the calculation with $\varphi=0$ and adjust the oscillation period $T_d=7.06(8)\,\mathrm s$ of the EC--decay of the H--like ion ${^{140}}{\rm Pr}^{58+}$. We get as a prediction of the squared mass difference $\Delta m^2_{21}=m^2_2-m^2_1=7.63(8)\,\cdot 10^{-5}\,\mathrm{eV}^2$. This value is in good agreement with the value $\Delta m^2_{21}=8.0^{+0.6}_{-0.5}\cdot 10^{-5}\,\mathrm{eV}^2$, obtained as a best--fit of the global analysis of the solar--neutrino and KamLAND experimental data \cite{Yao:2006px} (see also \cite{Fogli:2006aa}). This agreement shows that we are on the right track with our picture of the kinematics.

It still remains to check whether the prediction depends on the angle $\varphi$, see Fig.~1, between $\kappa$ and $k_l$. It turns out that for given $\Delta m^2_{21}$ the value of the energy violation $\Delta E$ is very stable against variations of $\varphi$ until close to $\pm\pi/2$ with an accuracy of at least five digits.  For $\varphi$ values larger than $\pi/2$ the solution starts to escape. This behaviour is understandable from the geometry and the kinematics. Due to the tiny difference $\vec\kappa$ of $\vec k_j$ and $\vec k_l$ the sphere with the radius $k_j$ is more a plain than a sphere. If $\varphi$ gets larger than $\pi/2$ the value of $\kappa$ increases abruptly and within a few thousandth of $1^\circ$ there is no solution of the equations possible. Furthermore, $\vec\kappa$ enters mainly as $\vec\kappa_\parallel$, as the component of $\vec\kappa$ parallel to $\vec k_l$. The component $\vec\kappa_\perp$, normal to $\vec k_l$ enters only by the term $\vec\kappa_\perp^2/(2M_d)$ which is negligible small.

Further, $\vec\kappa_\perp^2/(2M_m)$ is a negligible small energy. Therefore the energy violation is almost completely ascribable to the daughter nucleus and can be approximated with good accuracy by
\begin{align}\label{ApproViola}
\Delta E_l(k,\kappa,\varphi)=\frac{(\vec k+\vec\kappa)^2-\vec k^2}{2M_d},
\end{align}
where $k_j$ and $k_l$ are solutions of Eq.~(\ref{EneCons}) for the neutrino mass eigenstates $m_j$ and $m_l$ and $\vec\kappa=\vec k_j-\vec k_l$. For given $\Delta m^2_{21}$ the absolute values of $m_j$ and $m_l$ have a negligible influence. There is some dependence on the $Q$--value, but this is relatively weak.

\section{Quantum field theoretical calculations}
After the discussion of the mechanism we can start to use standard quantum field theory to verify the applicability of this mechanism. According to the standard theory of weak interactions one can describe the electron capture by the Hamiltonian
\begin{align}\label{weakH} 
 &\hspace{-0.3mm}H_W(t)=\frac{G_FV_{ud}}{\sqrt{2}}\int\mathrm d^3x\,
 [\bar\psi_n(x)\gamma^{\mu}(1-g_A\gamma_5)\psi_p(x)]\nonumber\\
 &\hspace{20mm}\times[\bar\psi_{\nu_e}(x)\gamma_{\mu}(1-\gamma_5)\psi_e(x)],
\end{align}
where $x = (t,\vec{r}\,)$, $G_F$ is the Fermi constant, $V_{ud}$ is a CKM matrix element, $g_A$ is the axial coupling constant \cite{Yao:2006px}, $\psi_n(x)$, $\psi_p(x)$, $\psi_e(x)$ and $\psi_{\nu_e}(x)$ are operators of the neutron, proton, electron and neutrino fields, respectively.

We take into account neutrino flavour mixing \cite{Yao:2006px}
\begin{align}\label{neutMix} 
\psi_{\nu_e}(x)=\sum_{j=1,2,3}U_{ej}^*\psi_{\nu_j}(x)
\end{align}
by the unitary neutrino--flavour mixing matrix $U$ with $U_{e1}=\cos\vartheta_{12}\cos\vartheta_{13}$, $U_{e2}=\sin\vartheta_{12}\cos\vartheta_{13}$ and $U_{e3}=\sin\vartheta_{13}\,\mathrm e^{-\mathrm i\delta_{CP}}$, where $\vartheta_{jl}$ are the mixing angles of the neutrino species $j$ and $l$. The phase $\delta_{CP}$ is a $CP$--violating phase \cite{Yao:2006px} which can be chosen as $\mathrm e^{\mathrm i\delta_{CP}}=\pm 1$. In our analysis the neutrinos $\nu_j\,(j=1,2,3)$ are Dirac particles with the masses $m_j$ and $\mathrm e^{\mathrm i\delta_{CP}}=1$.

First we evaluate the matrix element
\begin{align}\label{elemSchematic} 
\langle f|H_W(t)|i_{M_F}\rangle
\end{align}
between the initial state~(\ref{iniSta}) of the H--like ion $^{140}\mathrm{Pr}^{58+}$ in the state with $I=1$, $F=\frac{1}{2}$ and $M_F=\pm\frac{1}{2}$ and vanishing centre of mass momentum and the final state
\begin{align}\label{finSta} 
|f\rangle=|^{140}\mathrm{Ce}^{58+}(\vec q)\,\nu_j(\vec k),t\rangle
\end{align}
of the nucleus $^{140}\mathrm{Ce}^{58+}$ in a $0^+$-state with momentum $\vec q$ and neutrino mass eigenstate $\nu_j$ with momentum $\vec k$ at time $t$. According to the discussion above we have to allow for the state of the mother ions has a more complicated structure then for the states of the daughter nuclei. We assume for the initial state $|i_{M_F}\rangle$ of the mother ions in the centre of mass system a wave--packet, centred with a width $\sigma_\kappa$ around momentum zero
\begin{eqnarray}\label{iniSta}
\begin{aligned}
|i_{M_F}\rangle=\int\frac{\mathrm d^3\kappa}{(4\pi^3\sigma_\kappa^2)^{3/4}}
&\mathrm e^{\textstyle\,-\frac{\vec\kappa^2}{2\sigma_\kappa^2}}\\
&\times|^{140}{\rm Pr}^{58+}(\vec\kappa)_{M_F},t\rangle,
\end{aligned}
\end{eqnarray}
where $M_F$ is the magnetic quantum number of the mother ion, with a total angular momentum $I_m=1$. A finite value of $\sigma_\kappa$ implies that the wave function of the mother nucleus is restricted to some region of the centre of mass coordinate $\vec R$. The wave function~(\ref{iniSta}) is normalised to one mother ion. This ansatz is justified by the energy uncertainty, introduced by the time differential detection of the mother and daughter ions with a time resolution shorter than the modulation period \cite{Litvinov:2008rk}. In the final states it is not necessary to introduce wave--packets \cite{Stodolsky:1998tc}. For the determination of the decay rate one has to integrate over all final states and there should be no difference whether the final states are counted as plane waves or as wave--packets.

It is useful to treat centre of mass motion, relative motion, spin wave functions and time dependence in the matrix element $\langle f|H_W(t)|i_{M_F}\rangle$ separately. In the centre of mass system of the wave--packet of the mother ion the wave functions depending on the centre of mass coordinate $\vec R$ read
\begin{align}\label{cenMasMot} 
\int\frac{\mathrm d^3\kappa}{V(4\pi^3\sigma_\kappa^2)^{3/4}}
\mathrm e^{\textstyle\,-\frac{\vec\kappa^2}{2\sigma_\kappa^2}}
\mathrm e^{\mathrm i(\vec\kappa-\vec q-\vec k)\vec R},
\end{align}
where the wave functions of neutrino and daughter nucleus are normalised to one particle in the volume $V$. Since the wave--packet of the mother ion is already normalised~(\ref{iniSta}), no further normalisation must be included.

Due to the point--like nature of the Fermi interaction the transformation of the electron in a neutrino and of a proton in a neutron must take place at the same relative coordinate $\vec r$ and results in the overlap of the spatial wave functions of electron, neutrino and the matrix element of the ``hadronic'' isospin operator $t_-^\mathrm{h}$ between initial and final nucleus
\begin{align}\label{relMot} 
&\langle^{140}\mathrm{Ce}^{58+} (\vec q)\,\nu_j(\vec k)|t_-^\mathrm{h}t_+^\mathrm{l}|^{140}\mathrm{Pr}^{58+}(\vec 0)\rangle=\nonumber\\
&\hspace{10mm}=:M_\mathrm{EC}.
\end{align}
The transition of the electron to the electron neutrino is taken into account with the weak (leptonic) isospin operator $t_+^\mathrm{l}$. Since the K-electron capture is an allowed transition the exponentials for the relative wave functions are substituted by unity and the matrix element $M_\mathrm{EC}$ does not depend on the momenta $\vec k$ and $\vec q$.

The spinorial part of the matrix element $\langle f|H_W(t)|i_{M_F}\rangle$ we evaluate in the non--relativistic limit for neutrons, protons, electrons and mass--less neutrino states, with the Dirac spinors
\begin{eqnarray}\label{spinors}
\begin{aligned}
&u_n=\binom{\varphi_n}{0},\quad &u_p=\binom{\varphi_p}{0},\\
&u_{\nu j}=\frac{1}{\sqrt{2}}\binom{\varphi_-}{-\varphi_-},\quad
&u_e=\binom{\varphi_e}{0}.
\end{aligned}
\end{eqnarray}
Here we have omitted additional normalisations which would finally cancel with the normalisations of the creation and annihilation operators and the phase space integrals and assume that the spin wave functions $\varphi$ are normalised to unity. Inserting these spinors in the Hamiltonian (\ref{weakH}) one gets within a few lines for the spinor part the well--known Fermi- and Gamov-Teller contributions
\begin{align}\label{FandGT} 
&[\bar u_n(x)\gamma^{\mu}(1-g_A\gamma_5)u_p(x)]
 [\bar u_{\nu j}(x)\gamma_{\mu}(1-\gamma_5)u_e(x)]\nonumber\\
&\hspace{-1.0mm}=\sqrt{2}[\langle\varphi_n\varphi_-|\varphi_p\varphi_e\rangle
 -g_A\langle\varphi_n\varphi_-|\vec\sigma^\mathrm{h}\vec\sigma^\mathrm{l}
 |\varphi_p\varphi_e\rangle],
\end{align}
where $\vec\sigma^\mathrm{h,l}$ acts in the hadronic and leptonic spin space, respectively.

The spin states $\vec I_m$ and $\vec I_d$ of mother and daughter nuclei we decompose in the spin $\vec I_c$ of a core and proton and neutron spins $\vec j_{p,n}$ and include angular momentum conservation by
\begin{align}\label{angMomCons} 
  \underbrace{\vec I_c+\vec j_n}_{\vec I_d=0}+\vec s_\nu=\vec F
 =\underbrace{\vec I_c+\vec j_p}_{\vec I_m}+\vec s_e,
\end{align}
where $\vec s_e$ and $\vec s_\nu$ are the spins of electron and neutrino.

To simplify the calculations we quantise the spins in the direction of the neutrino momentum $-\vec q$. Therefore, the magnetic quantum number of the neutrino state (\ref{spinors}) is $\mu_\nu=-1/2$ and due to the vanishing spin of the daughter nuclei $I_d=0$ we get $M_F=-1/2$ and  $I_c=1/2$. In the initial state and in the final state we have to couple three spin-1/2-states to $F=1$ and $M_F=-1/2$. Since the mother nucleus has $I_m=1$, the initial spin state $|\phi_i\rangle$ is a mixed symmetric doublet state, it reads in the representation with magnetic quantum numbers $|\mu_c\mu_p\mu_e\rangle$
\begin{align}\label{spinIni} 
|\phi_i\rangle=\frac{1}{\sqrt{6}}
|\uparrow\downarrow\downarrow+\downarrow\uparrow\downarrow
 -2\downarrow\downarrow\uparrow\rangle.
\end{align}
Due to $I_d=0$ the final spin state $|\phi_f\rangle$ is a mixed antisymmetric doublet state, it reads in the representation $|\mu_c\mu_n\mu_\nu\rangle$
\begin{align}\label{spinFin} 
|\phi_f\rangle=\frac{1}{\sqrt{2}}
|\uparrow\downarrow\downarrow-\downarrow\uparrow\downarrow\rangle.
\end{align}
Since $|\phi_i\rangle$ and $|\phi_f\rangle$ are orthogonal states, the Fermi contribution in (\ref{FandGT}) vanishes. Inserting these states in the GT-contribution in (\ref{FandGT}) the spin operators $\vec\sigma^\mathrm{h,l}$ leave the spin wave function of the core untouched, and the spinor part of the matrix element $\langle f|H_W(t)|i_{M_F}\rangle$ reads after a short calculation
\begin{align}\label{GTspinPart} 
&-\sqrt{2}g_A\langle\varphi_n\varphi_-|\vec\sigma^\mathrm{h}\vec\sigma^\mathrm{l}
 |\varphi_p\varphi_e\rangle]=-\sqrt{6}\,g_A.
\end{align}

We collect now all the terms contributing to the matrix element $\langle f|H_W(t)|i_{M_F}\rangle$, the spinor contribution (\ref{GTspinPart}), the prefactor $G_FV_{ud}/\sqrt{2}$ in the Hamiltonian (\ref{weakH}), the integral over the spatial wave functions (\ref{relMot}), the integral over the centre of mass coordinate (\ref{cenMasMot}), the neutrino mixing (\ref{neutMix}) in the wave function, the exponential for the time--dependence according to Eq.~(\ref{OffShellE}) and get with the abbreviation
\begin{align}\label{GTElem}
M_\mathrm{GT}=-\sqrt{3}g_AG_FV_{ud}M_\mathrm{EC}
\end{align}
the matrix element
\end{multicols}
\rule{64mm}{.1pt}
\begin{align}\label{MElement}
\hspace{-2.5mm}\langle f|H_W(t)|i_{M_F}\rangle
=\delta_{M_F,-\frac{1}{2}}M_\mathrm{GT}
\int\frac{\mathrm d^3\kappa}{V(4\pi^3\sigma_\kappa^2)^{3/4}}
\mathrm e^{\textstyle\,-\frac{\vec\kappa^2}{2\sigma_\kappa^2}}
\int\mathrm d^3R\mathrm\;\mathrm e^{\mathrm i(\vec\kappa-\vec q-\vec k)\vec R}
\sum_{j=1,2,3}U_{ej}\exp\{\mathrm i\Delta E_jt\}.
\end{align}
\hspace{90mm}\rule{64mm}{.1pt}
\begin{multicols}{2}
\noindent For the further calculation it is important to evaluate the $\vec R$ integral carefully for finite extents $L$, or $L_i$, of the volume $V$ and not just to substitute the integral by a delta--function. In Eq.~(\ref{MElement}) we have also taken into account that for $M_F=1/2$ there is no transition possible.

Since the interaction is weak it is sufficient to use time dependent perturbation theory of first order to determine the decay rate for the electron capture reaction. We average over the two initial states $|i_{M_F}\rangle$ with $M_F=\pm 1/2$, integrate the square of the time--integrated matrix element~(\ref{MElement}) over the possible final states with the density $(2\pi)^3 /V$ in momentum space and get for the decay factor
\begin{align}\label{decaRate}
\lambda_\mathrm{EC}(t)=&\frac{1}{2}\sum_{M_F}\frac{\mathrm d}{\mathrm dt}
\int\frac{\mathrm d^3k}{(2\pi)^3/V}\int\frac{\mathrm d^3q}{(2\pi)^3/V}\nonumber\\
&\times\Big|\int_0^t\mathrm d\tau\langle f|H_W(\tau)|i_{M_F}\rangle\Big|^2.
\end{align}
Here we have inserted an important modification of the usual expression which will turn out to be necessary for the further calculation. The decay rate is derived from the percentage of decayed nuclei by a time derivative and not by a division by the time $t$, as usual.

Inserting the matrix element (\ref{MElement}) we can easily simplify the expression~(\ref{decaRate}) for the decay factor. We sum over $M_F$, integrate the coordinate $\vec R$ over the volume $V=L^3$ and cancel the volume factors.

For the  evaluation of the $\vec q$--integral over the product of strongly oscillating $\sin$--functions in the decay factor~(\ref{decaRate}) we will use a formula, similiar to the well--known limit
\begin{align}\label{intSinSq} 
\hspace{-3mm}\int\mathrm d k\frac{\sin^2(kt)}{k^2}f(k)
\to\pi t\int\mathrm d k f(k)\delta(k)
\end{align}
which is valid for large $t$ and for smooth functions $f(k)$. The limit
\begin{align}\label{intSinSin}
&\hspace{-3mm}\int\mathrm dk
\frac{\sin[(k-k_1)t]}{k-k_1}\frac{\sin[(k-k_2)t]}{k-k_2}f(k)
\stackrel{|k_1-k_2|t\gg 2\pi}{\longrightarrow}\nonumber\\
&\to\frac{\pi}{2}\int\mathrm dk\Big\{\delta(k-k_1)\frac{\sin[(k-k_2)t]}{k-k_2}+\\
&\hspace{15mm}+\frac{\sin[(k-k_1)t]}{k-k_1}\delta(k-k_2)\Big\}f(k).\nonumber
\end{align}
is valid for $|k_1-k_2|t\gg 2\pi$ and is discussed in the appendix. For $\sigma_\kappa\gg 1/L$ we can assume that the Gaussians are smooth functions. With these calculations we get
\end{multicols}
\rule{64mm}{.1pt}
\begin{eqnarray}\label{decaRateCalc1}
\begin{aligned}
\lambda_\mathrm{EC}(t)&=\frac{M_\mathrm{GT}^2}{2}
\int\frac{\mathrm d^3k}{(2\pi)^3}
\int\frac{\mathrm d^3\kappa}{(4\pi^3\sigma_\kappa^2)^{3/2}}
\prod_{i=1}^3\Big[\int\mathrm d\kappa_i^\prime
\frac{\sin(\frac{\kappa_i-\kappa_i^\prime}{2}L)}{\frac{\kappa_i-\kappa_i^\prime}{2}}
\,\mathrm e^{\textstyle\,-\frac{\kappa_i^2+\kappa_i^{\prime 2}}{2\sigma_\kappa^2}}\Big]\\
&\times\frac{\mathrm d}{\mathrm dt}\Big[\sum_jU_{ej}\int_0^t\mathrm d\tau
\exp\{\mathrm i\Delta E_j(k,\kappa,\varphi)t\}\Big]
\Big[\sum_lU_{el}^*\int_0^t\mathrm d\tau
\exp\{-\mathrm i\Delta E_lt(k,\kappa^\prime,\varphi^\prime)\}\Big]
\end{aligned}
\end{eqnarray}
\hspace{90mm}\rule{64mm}{.1pt}
\begin{multicols}{2}

The time integrals in Eq.~(\ref{decaRateCalc1}) integrate over the off-shell contributions~(\ref{OffShellE})
\begin{align}\label{timInt} 
\int_0^t\mathrm d\tau\exp\{\mathrm i\Delta E_j\tau\}=
\frac{\sin\frac{\Delta E_jt}{2}}{\frac{\Delta E_j}{2}}\,
\mathrm e^{\mathrm i\,\frac{\Delta E_j}{2} t}.
\end{align}
The sum over the flavour contributions $j$ in Eq.~(\ref{decaRateCalc1}) has to be squared and contains interference terms. These terms oscillate in time with constant amplitude. A division by $t$, as it is usually applied in the determination of the decay rate from the square of the amplitude, would destroy them. We use therefore the correct definition of the rate as the derivative of the square of the amplitude and get for real flavour mixing coefficients $U_{ej}$:
\begin{align}\label{timDer} 
&\frac{\mathrm d}{\mathrm dt}
\Big|\sum_jU_{ej}\frac{\sin\frac{\Delta E_jt}{2}}{\frac{\Delta E_j}{2}}\,
\mathrm e^{\mathrm i\,\frac{\Delta E_j}{2} t}\Big|^2=\\
&\hspace{2mm}=2\pi\sum_j\delta(\Delta E_j)\Big[U_{ej}^2
+\sum_{l(\ne j)}U_{ej}U_{el}\cos(\Delta E_lt)\Big]\nonumber
\end{align}
with interference terms which survive for large $t$. The energy conserving delta--function $\delta(\Delta E_j)$ has a negligible dependence on $m_j$
\begin{align}\label{DerOffShell}
\frac{\partial \Delta E_j(k,\kappa,\varphi)}{\partial m_j}=\frac{m_j}{Q}
\end{align}
and can be moved out of the sum over the neutrino species, $\sum_j\delta(\Delta E_j)\rightarrow\delta(\Delta E)\sum_j$. Due to the smallness of the recoil of the daughter nucleus the derivatives of $\Delta E(k,\kappa,\varphi)$ with respect to $k$ and $\kappa$ are unmeasurable close to unity. From Eq.~(\ref{OffShellE}) we get
\begin{align}\label{OffShellDer}
&\frac{\partial \Delta E_j(k,\kappa,\varphi)}{\partial k}=1+\frac{Q}{M_d}
=1.000026,\nonumber\\
&\frac{\partial \Delta E_j(k,\kappa,\varphi)}{\partial \kappa}=1-\frac{Q}{M_d}
=0.999974.
\end{align}
Therefore, $\delta(\Delta E)$ selects the modulus $k$ of the neutrino momentum in the expression~(\ref{decaRateCalc1}) for the decay factor and modifies its value by some ppm only.

For the further calculations it is simpler to treat diagonal and off--diagonal terms separately. In the diagonal terms of Eq.~(\ref{decaRateCalc1}) one can integrate the sin function over $\kappa_i^\prime$ enforcing $\kappa_i=\kappa_i^\prime$. With the help of Eq.~(\ref{timDer}) and using the unitarity of the flavour mixing matrix we can drastically simplify the contribution of the diagonal terms to the decay amplitude
\begin{align}\label{decaRateAve}
\bar\lambda&_\mathrm{EC}=\frac{M_\mathrm{GT}^2}{2}\int\frac{\mathrm d^3k}{(2\pi)^2}\\
&\times\int\frac{\mathrm d^3\kappa}{(\pi\sigma_\kappa^2)^{3/2}}\,
\,\mathrm e^{\textstyle\,-\frac{\vec\kappa^2}{\sigma_\kappa^2}}
\delta(\Delta E(k,\kappa,\varphi)).\nonumber
\end{align}
Small variations of the centre of mass momentum lead to a tiny variation of the momenta of the emitted neutrinos. Their contributions to the averaged decay factor $\bar\lambda_\mathrm{EC}$ can be summed up by an integration over the momentum $\vec\kappa$ of the wave--packet of the mother nucleus. The energy conservation $\delta(\Delta E)$ restricts the momenta $\vec k$ of the neutrinos to $k=Q$. Therefore we get the average electron capture constant
\begin{align}\label{decaRateAveFin}
\bar\lambda_\mathrm{EC}&=\frac{M_\mathrm{GT}^2}{2}
\int\frac{\mathrm d^3k}{(2\pi)^2}\delta(\Delta E)
=\frac{M_\mathrm{GT}^2Q^2}{2\pi}=\nonumber\\
&=\frac{3}{2\pi}[g_AG_FV_{ud}M_\mathrm{EC}Q]^2.
\end{align}

The terms in Eq.~(\ref{decaRateCalc1}), which are non--diagonal in the neutrino flavours, describe the interference of the mass eigenstates $m_j$ and $m_l$ of the neutrinos with momentum $\vec k$ and lead to an interesting oscillation of the decay factor with the frequency $\Omega$ given by the energy violation $\Omega=\Delta E$. With Eq.~(\ref{timDer}) such a term reads
\end{multicols}
\rule{64mm}{.1pt}
\begin{eqnarray}\label{decaRateOszFin}
\begin{aligned}
\lambda_\mathrm{EC}^{jl}(t)&=\frac{M_\mathrm{GT}^2}{2}\int\frac{\mathrm d^3k}{(2\pi)^3}
\prod_{i=1}^3\Big[\frac{\int\mathrm d\kappa_i\int\mathrm d\kappa_i^\prime}
{(4\pi^3\sigma_\kappa^2)^{1/2}}
\,\mathrm e^{\textstyle\,-\frac{\kappa_i^2+\kappa_i^{\prime 2}}{2\sigma_\kappa^2}}
\frac{\sin(\frac{\kappa_i-\kappa_i^\prime}{2}L)}
{\frac{\kappa_i-\kappa_i^\prime}{2}}\Big]\\
&\hspace{5mm}\times U_{ej}U_{el}\frac{\mathrm d}{\mathrm dt}\int_0^t\mathrm d\tau
\exp\{\mathrm i\Delta E_j(k,\kappa^\prime,\varphi^\prime)t\}
\int_0^t\mathrm d\tau
\exp\{-\mathrm i\Delta E_lt(k,\kappa,\varphi)\}=\\
&=\frac{M_\mathrm{GT}^2}{2}\int\frac{\mathrm d^3k}{(2\pi)^3}
\prod_{i=1}^3\Big[\int\frac{\mathrm d\kappa_i}{2\pi}
\frac{\sin(\frac{\kappa_i}{2}L)}{\frac{\kappa_i}{2}}
\,\mathrm e^{\textstyle\,-\frac{\kappa_i^2}{4\sigma_\kappa^2}}\Big]
U_{ej}U_{el}2\pi\delta(\Delta E_j(k,0,\varphi))
\cos(\Delta E_l(k,\kappa,\varphi)t)
\end{aligned}
\end{eqnarray}
\hspace{90mm}\rule{64mm}{.1pt}
\begin{multicols}{2}
\noindent Here we have used that the integrand is a function of $\vec\kappa-\vec\kappa^\prime$ only and that folding two normalised Gaussian distributions of width $\sigma$ gives a normalised Gaussian of width $\sigma\sqrt{2}$. In Eq.~(\ref{decaRateOszFin}) both neutrinos, $m_j$ and $m_l$, are integrated over the same momenta $\vec k$. The neutrino $m_j$ is on--shell $\vec k_j=\vec k$  and the emission of the neutrino $m_l$ due to the neutrino momentum $\vec k_l=\vec k-\vec\kappa$ in the centre of mass system is slightly off--shell by $\Delta E_l(k,\kappa,\varphi)$ according to Eq.~(\ref{OffShellE}). The value of the energy violation $\Delta E_l(k,\kappa,\varphi)$ defines the oscillation frequency in Eq.~(\ref{decaRateOszFin}). It's the energy violating contributions to the decay rate which lead to measurable consequences, to the modulation of the decay rate in time.

\section{Discussion and concluding remarks}
There are two restrictions for the region of allowed $\vec\kappa$ values, the width $\sigma_\kappa$ of the wave--packet and the spatial extent $L$ of the wave function of the mother ion. If the form of the wave--packet is induced by the spatial extent $L$ of the system then there should be a relation of the form $\sigma_\kappa=2\pi/L$. In the limit $L\to\infty$ the sin-function in Eq.~(\ref{decaRateOszFin}) leads to a delta--function $\delta(\kappa_i)$. The vanishing momentum $\vec\kappa$ does not allow to solve the kinematic equation (\ref{OffShellE}) as can also be seen in Fig.~1 where the shaded sphere of allowed $\vec\kappa$ values shrinks to zero and the area $A$ of the overlap with the sphere defined by the momentum $k=k_j$ gets also zero. As a consequence there are no oscillations. There are also no oscillations if $\sigma_\kappa$ is very small. Only if the allowed momenta $\kappa$ are larger than $|k_j-k_l|$ the sphere with radius $k=k_j$ crosses the $\kappa$ cloud in a finite area $A$ which we can estimate with the minimum of $\sigma_\kappa^2-(k_1-k_2)^2$ and $(2\pi/L)^2-(k_1-k_2)^2$,
\begin{align}\label{area}
&A(|k_1-k_2|,L,\sigma_\kappa)=\\
&=\mathrm{Min}\{\sigma_\kappa^2-(k_1-k_2)^2,(2\pi/L)^2-(k_1-k_2)^2\}\nonumber
\end{align}
Here we have assumed for simplicity spherical symmetry. For an estimate of the oscillation amplitude we have to multiply $A$ with the integrals over $\kappa_i$ in the square brackets on the l.h.s. of Eq.~(\ref{decaRateOszFin}). The product of these three integrals is a function of $F_3(L\cdot\sigma_\kappa)$. This function
\begin{align}\label{funF}
F_n(L\cdot\sigma_\kappa)=\prod_{i=1}^n\Big[\int\frac{\mathrm d\kappa_i}{2\pi}
\frac{\sin(\frac{\kappa_i}{2}L)}{\frac{\kappa_i}{2}}
\,\mathrm e^{\textstyle\,-\frac{\kappa_i^2}{4\sigma_\kappa^2}}\Big]
\end{align}
is plotted in Fig.~2.
\end{multicols}
\rule{64mm}{.1pt}
\begin{figure}[h]
\psfrag{k1}{$\vec k_j$}\psfrag{k2}{$\vec k_l$}\psfrag{f}{$\varphi$}\psfrag{k}{$\vec\kappa$}
\centering
\includegraphics[scale=0.75]{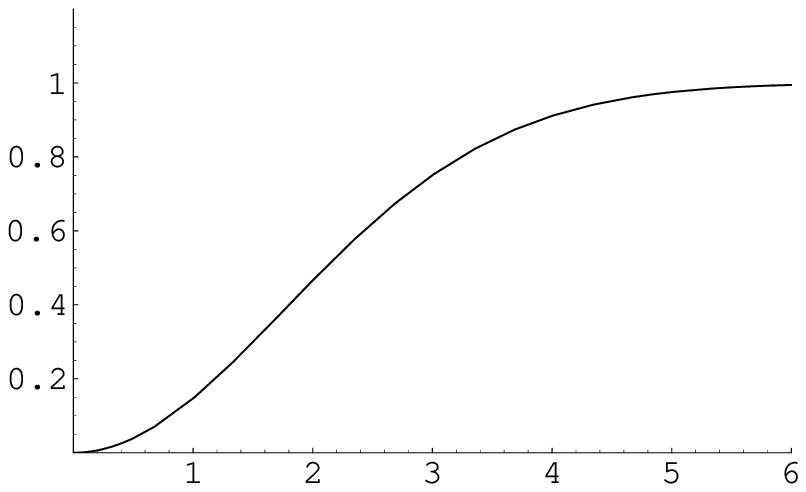}
\includegraphics[scale=0.75]{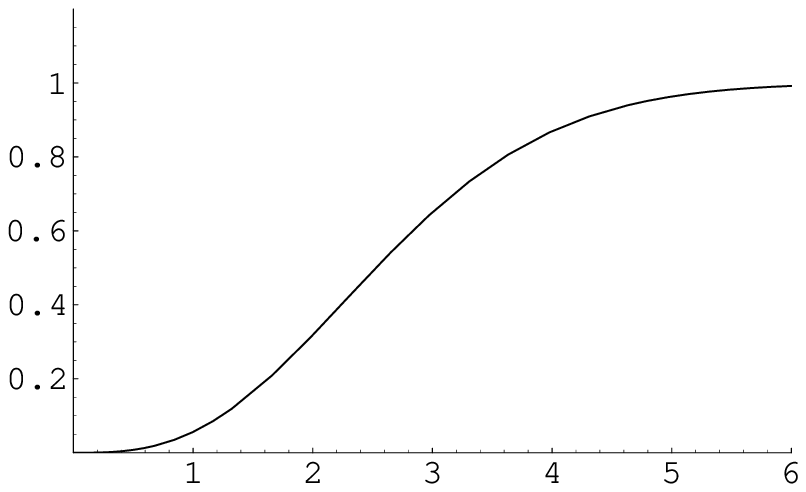}
\caption{Function $F_2(L\cdot\sigma_\kappa)$ (left) and function $F_3(L\cdot\sigma_\kappa)$ (right), as defined in Eq.~(\ref{funF}).}
\label{fun3}
\end{figure}
\begin{multicols}{2}

$\lambda_\mathrm{EC}^{jl}(t)$ differs from $\bar\lambda_\mathrm{EC}$ by the factors $A F _3U_{ej}U_{el}\cos(\Delta E_lt)$. Under the assumption that the two lightest neutrino flavours contribute only we have $U_{e1}U_{e2}=1/2\cdot\sin(2\vartheta_{12})$. We get therefore a relation for the oscillation amplitude,
\begin{align}\label{amplitude}
a=\frac{A(|k_1-k_2|,L,\sigma_\kappa) F_3(L\cdot\sigma_\kappa)}{2}\sin(2\vartheta_{12}).
\end{align}

For simplicity we have assumed that the volume $V$ has the same extent in 3 directions. But most likely due to the beam geometry a cylindrical symmetry is more likely with $L_x\approx L_y$ and $L_z=\infty$. This would mean that the neutrinos can interfere only if they are emitted in a direction perpendicular to the beam. The product of integrals should then be extended over the $x$ and $y$ directions only and the function $F_2$ should be used instead of $F_3$.

For the experimental value of the flavour mixing angle $\sin(2\vartheta_{12})=0.925(32)$~\cite{Yao:2006px},\cite{Fogli:2006aa} and the experimental value\cite{Litvinov:2008rk} of the amplitude $a=0.15(3)$ we get from Eq.~(\ref{amplitude}) for the factor $A\cdot F=0.32$.

There remains the question why the value for the oscillation frequency is so close to the value expected from the experimental value of the squared mass difference $\Delta m^2_{21}=m^2_2-m^2_1=(0.763\pm 8)\,\cdot 10^{-4}\,\mathrm{eV}^2$~\cite{Yao:2006px}. It seems obvious that a Lorentz factor for the transformation from the centre of mass system to the laboratory system would be necessary. This factor should be $\gamma=1.43$ and would increase the oscillation period. Since no essential approximation was made in the quantum field theoretical determination of this period one has to think about the possible reasons for this fact. It seems that we have to take into account that the ions don't move in an inertial system but around a circle and that the emission time of the neutrino takes longer then the rotation period. How could the neutrinos lead to oscillations of such a long period if the emission process would be shorter than the revolution period.

Finally, we would like to give a recipe to predict oscillation frequencies in other EC-capture systems. First one solves Eq.~(\ref{EneCons}) to get $k_1$ and $k_2$. The mass $m_j$ of the lighter neutrino is not important, as long as it is in a reasonable region, since $k_1-k_2$ is not influenced by this value. The difference of squared masses is of course very crucial. From  $\vec k_1$ and $\vec k_2$ one can determine $\vec\kappa=\vec k_1-\vec k_2$. Then the oscillation frequency due to the interference of the mass eigenstates $m_1$ and $m_2$ is given by $\Delta E[k_1,\kappa,\cos\varphi,m_2]$ as defined in Eq.~(\ref{OffShellE}). $\Delta E[k_1,0,\cos\varphi, m_1]$ is by definition zero. The angle $\varphi$ can be put to zero, its absolute value should not be larger or very close to $\pi/2$. As already mentioned this calculation has to be done with sufficient accuracy. At least 30 digits are recommended. Too large errors have catastrophic consequences.

Detailed numbers for the EC--capture decay of ${^{140}}{\rm Pr}^{58+}$ are shown in Table~\ref{tab1}. For the second EC--capture reaction which was published in~\cite{Litvinov:2008rk}, the electron capture decay of the H--like ion ${^{142}}{\rm Pm}^{60+}$ to $^{142}{\rm Nd}^{60+}$ we get a prediction for $\Delta m^2_{21}=0.775(26)\,\cdot 10^{-4}\,\mathrm{eV}^2$. The corresponding numbers are listed in table~\ref{tab2}.

In conclusion, we are able to describe quantitatively the periodic modulations of the expected exponential decrease of the number of EC--decays per time which were observed in recent electron capture experiments in the Experimental Storage Ring (ESR) at GSI in Darmstadt \cite{Litvinov:2008rk} by neutrino mixing. We give general formulae which relate the modulation frequency with the difference $\Delta m^2_{jl}$ of squared masses of two neutrino mass eigenstates $m_j$ and $m_l$. With these formulae we get for both reactions where such modulations were observed, the nuclear K--shell electron captures of the H--like ions ${^{140}}{\rm Pr}^{58+}$ and ${^{142}}{\rm Pm}^{60+}$, values of $\Delta m^2_{21}$ which agree nicely with the global analysis of the solar--neutrino and KamLAND experimental data \cite{Yao:2006px}. These general formulae which allow also to predict the modulation frequencies in further EC and bound beta decay experiments depend only on kinematical parameters of the process. We discuss that for the appearence of these modulations it is necessary to have experiments which produce a sufficient momentum uncertainty for the wave--functions (wave--packets) of the mother ions. The amplitude of the modulations is determined by the widths of the wave--packets, but the frequency depends on $\Delta m^2_{21}$ only.

\vspace{3mm}
When this article was in the final stage of its preparation we realised that H.~J.~Lipkin has also published an article about the problem of neutrino mixing in K-capture~\cite{Lipkin:2008ai}. In~\cite{Ivanov:2008sd} the same GSI--experiment~\cite{Litvinov:2008rk} was explained by neutrino mixing using an ansatz for wave--packets in the final state only.

\section{Acknowledgement}
I would like to thank especially to Paul Kienle for the formulation of this interesting question, for many intensive discussions and for his strong believe that the origin of the oscillations of the decay factor is due to neutrino mixing. I thank to Andrei Ivanov for his critical comments which lead to a deeper understanding of the problem and to Alexander Kobushkin for a critical reading of the manuscript and for his suggestions to improve the readability. Further I am grateful to Max Kreuzer for an interesting discussion about the duration of an emission process.

\end{multicols}
\begin{table}
\begin{center}
\begin{tabular}{|c|r@{.}l|}
\hline
$M_d$&   130319252000&000000000000000000000000000000000000000\\
$Q$&          3388000&000000000000000000000000000000000000000\\
\hline
$m_1$&              0&100000000000000005551115123125782702118\\
$m_2$&              0&100380775051799640446407847251541936522\\
$\Delta m_{21}^2$&   0&000076299999999999997923709471603359588\\
\hline
$k_1$&        3387955&961051731594435668834899140341414454486\\
$k_2$&        3387955&961051731583175484621991824131728111711\\
$\kappa$&           0&000000000011260184212907316209686342774\\
\hline
$E\nu_1$&     3387955&961051733070251782586223414251214280672\\
$E\nu_2$&     3387955&961051733070252075321238703166027210199\\
$E\nu_2^\prime$&3387955&961051733070252075321238703166027210199\\
\hline
$E_{d1}$&            44&038948266929748217413776585748785719327\\
$E_{d2}$&            44&038948266929747924678761296833972789800\\
$E_{d2}^\prime$&       44&038948266929748510148791874663599621787\\
$E_{d2}^\prime-E_{d2}$&  0&000000000000000585470030577829626831986\\
$E_{m2}^\prime$&        0&000000000000000000000000000000000486453\\
$\Delta E$&          0&000000000000000585470030577829626345533\\
\hline
\end{tabular}
\end{center}
\caption{Some quantities in eV, relevant for the explanation of the oscillation of the decay factor of the H--like ion ${^{140}}{\rm Pr}^{58+}$ to $^{140}{\rm Ce}^{58+}$. The calculations are done for $\varphi=0$ according to Eq.~(\ref{EneCons}) and Eq.~(\ref{OffShellE}). $m_1$ is on--shell and $m_2$ off--shell. $E_{\nu_2}$ and $E_{d2}$ are the energies of $m_2$ and the daughter nucleus when the neutrino is emitted from the mother nucleus at rest. $E_{\nu_2}^\prime$ and $E_{d2}^\prime$ are the corresponding energies when $m_2$ is emitted from a mother nucleus moving with momentum $-\kappa$ against $k_1$ in order to get $k_2=k_1-\kappa$.}
\label{tab1}
\end{table}

\begin{table}
\begin{center}
\begin{tabular}{|c|r@{.}l|}
\hline
$M_d$&   132321536600&000000000000000000000000000000000000000\\
$Q$&          4870000&000000000000000000000000000000000000000\\
\hline
$m_1$&              0&100000000000000005551115123125782702118\\
$m_2$&              0&100386752114011543561103471770365429653\\
$\Delta m_{21}^2$&   0&000077499999999999999986989573930173946\\
\hline
$k_1$&        4869910&384855836089081365647620981219735553232\\
$k_2$&        4869910&384855836081124633213067424714490018121\\
$\kappa$&           0&000000000007956732434553556505245535110\\
\hline
$E\nu_1$&     4869910&384855837115794303846383129612591300565\\
$E\nu_2$&     4869910&384855837115794596682869979717867103615\\
$E\nu_2^\prime$&4869910&384855837115794596682869979717867103615\\
\hline
$E_{d1}$&            89&615144162884205696153616870387408699434\\
$E_{d2}$&            89&615144162884205403317130020282132896384\\
$E_{d2}^\prime$&       89&615144162884205988990103720492684980937\\
$E_{d2}^\prime-E_{d2}$&  0&000000000000000585672973700210552084552\\
$E_{m2}^\prime$&        0&000000000000000000000000000000000239217\\
$\Delta E$&          0&000000000000000585672973700210552084552\\
\hline
\end{tabular}
\end{center}
\caption{Same as Table~\ref{tab1} for the electron capture decay of the H--like ion ${^{142}}{\rm Pm}^{60+}$ to $^{142}{\rm Nd}^{60+}$.}
\label{tab2}
\end{table}

\clearpage
\section{Appendix}\begin{itemize}
\item Comments to Eq.~(\ref{intSinSin})\\
\begin{enumerate}
\item We define the functions
\begin{align}\label{ssDef} 
&ss(x,t)=\frac{\sin[(x-x_0)t]\sin[(x+x_0)t]}{x^2-x_0^2}=\frac{\cos(2x_0t)-\cos(2xt)}{2(x^2-x_0^2)}\\
&i(X,t)=\int_{-X}^{+X}\mathrm dx ss(x,t)\\
&I(t)=\int_{-\infty}^{+\infty}\mathrm dx ss(x,t)
\end{align}
and get the analytical relation
\begin{eqnarray}\label{ssInt} 
\begin{aligned}
i(X,t)=\int_{-X}^X\mathrm dx ss(x,t)&
=\int\mathrm dx\frac{\sin[(x-x_0)t]\sin[(x+x_0)t]}{x^2-x_0^2}
=\int\mathrm dx\frac{\cos(2x_0t)-\cos(2xt)}{2(x^2-x_0^2)}=\\
=\frac{2}{x_0}\Big\{&\sin(2tx_0)
\big[\int_0^{2t(X-x_0)}\mathrm dx\frac{\sin x}{x}
+\int_0^{2t(X+x_0)}\mathrm dx\frac{\sin x}{x}\big]\\
+&\cos(2tx_0)\big[\int_{2t(X-x_0)}^\infty\mathrm dx\frac{\cos x}{x}
-\int_{2t(X+x_0)}^\infty\mathrm dx\frac{\cos x}{x}-2\mathrm{atanh}\frac{X}{x_0}\big]
\Big\}.
\end{aligned}
\end{eqnarray}

\item $I(t)$ is periodic in $t$ with $\omega=2x_0$ and takes its maximal value at $2tx_0=2n\pi+\frac{\pi}{2}$. At such $t$--values the expression~(\ref{ssInt}) simplifies to
\begin{align}
i(X,t)=\frac{2}{x_0}\Big\{\int_0^{2t(X-x_0)}\mathrm dx\frac{\sin x}{x}
+\int_0^{2t(X+x_0)}\mathrm dx\frac{\sin x}{x}\Big\}
\end{align}
The following diagram depicts $i(X,t)$ at $t=8\pi+\frac{\pi}{4}$ for $x_0=1$. One can clearly see the influence of the two peaks of $ss(x,t)$ at $\pm x_0$.\\
\includegraphics[scale=0.8]{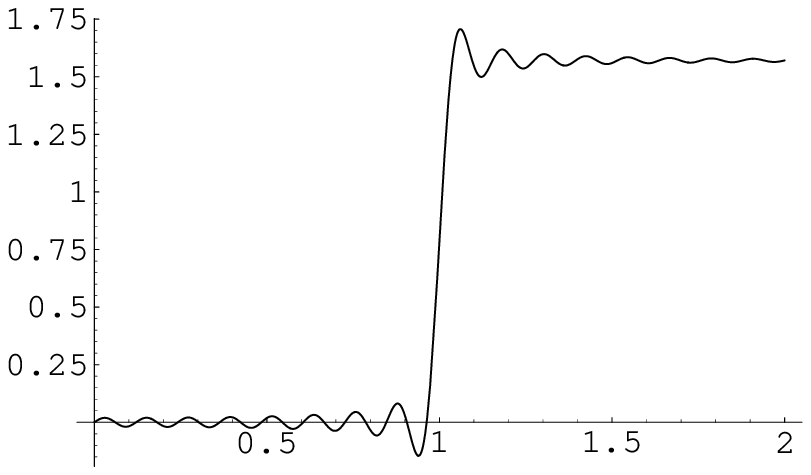}\\
The values at these peaks $ss(\pm x_0,t)$ oscillate with $t$ and with an amplitude increasing proportional to $t$
\begin{align}\label{limits} 
ss(\pm x_0,t)=\frac{t\sin(2x_0t)}{2x_0}
\end{align}
The widths of these peaks of $ss(X,t)$ at $X=\pm x_0$ and $t=(n\pm\frac{1}{4})\frac{\pi}{x_0}$ shrink proportional to $1/t$.

\item $I(t)$ is oscillating with the same frequency like $ss(\pm x_0,t)$.\\
$I(t)$ has a zero at $t=n\frac{\pi}{x_0}$. The following diagrams shows
 $ss(X,t)$ for $t=8\frac{\pi}{x_0}$.\\
\includegraphics[scale=0.8]{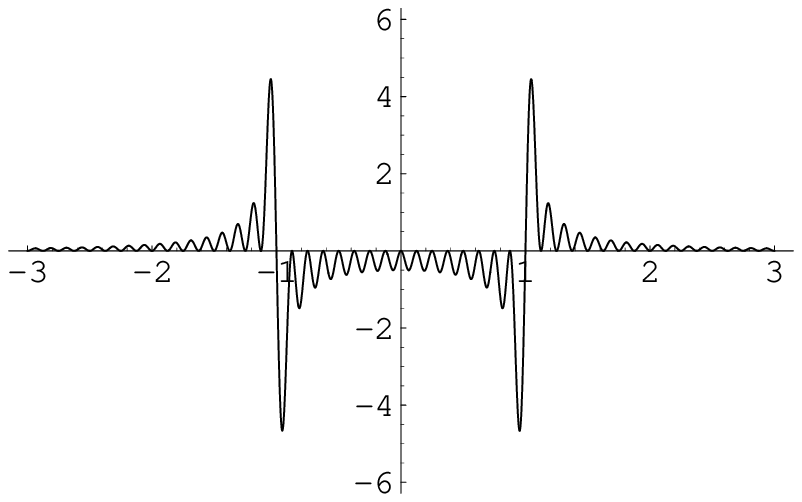}\\
$I(t)$ has a maximum at $t=(n+\frac{1}{4})\frac{\pi}{x_0}$. The following diagram
 shows $ss(X,t)$ for $t=(8+\frac{1}{4})\frac{\pi}{x_0}$.\\
\includegraphics[scale=0.8]{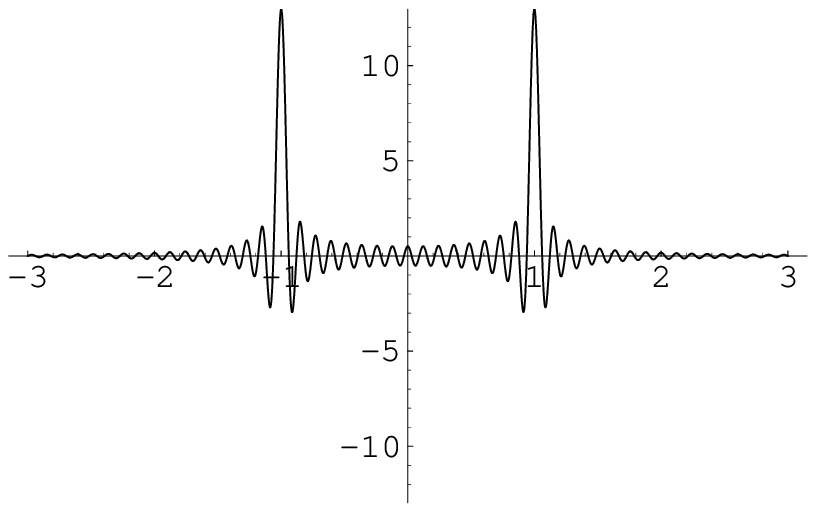}\\
$I(t)$  has a zero at $t=(n+\frac{1}{2})\frac{\pi}{x_0}$. The following diagram shows
 $ss(X,t)$ for $t=(8+\frac{1}{2})\frac{\pi}{x_0}$.\\
\includegraphics[scale=0.8]{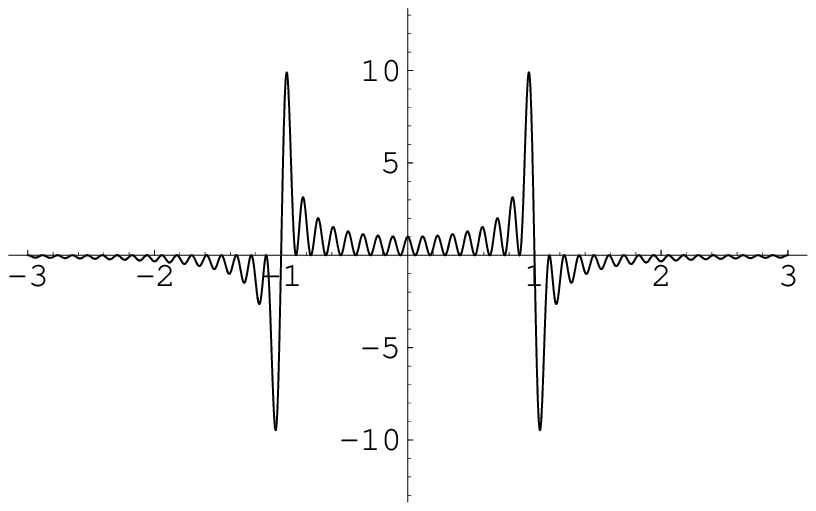}\\
$I(t)$  has a minimum at $t=(n+\frac{3}{4})\frac{\pi}{x_0}$. The following diagram shows
 $ss(X,t)$ for $t=(8+\frac{3}{4})\frac{\pi}{x_0}$.\\
\includegraphics[scale=0.8]{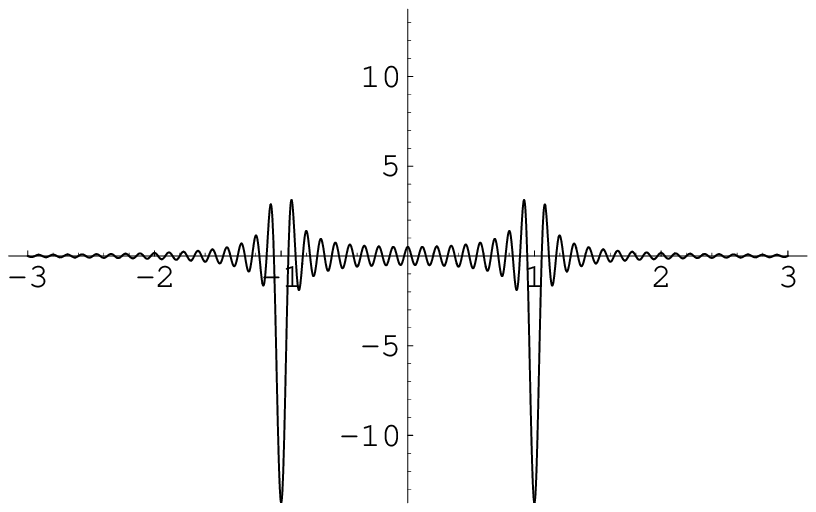}\\

\end{enumerate}

\item Some modification of Eq.~(\ref{intSinSin})\\
\begin{align}\label{intSinSinHalf}
&\hspace{-3mm}\int\mathrm dk
\frac{\sin[(k-k_1)L/2]}{(k-k_1)/2}\frac{\sin[(k-k_2)L/2]}{(k-k_2)/2}f(k)
\stackrel{|k_1-k_2|L/2\gg 2\pi}{\longrightarrow}\nonumber\\
&\to\pi\int\mathrm dk\Big\{\delta(k-k_1)
\frac{\sin[(k-k_2)L/2]}{(k-k_2)/2}
+\frac{\sin[(k-k_1)L/2]}{(k-k_1)/2}\delta(k-k_2)\Big\}f(k)=\nonumber\\
&=\pi\frac{\sin[(k_1-k_2)L/2]}{(k_1-k_2)/2}[f(k_1)+f(k_2)]
\end{align}

\item Detailed derivation of Eq.~(\ref{decaRateCalc1}) from Eq.~(\ref{decaRate})
\begin{align}\label{decaRatek1} 
&\lambda_\mathrm{EC}(t)
=\frac{1}{2}\frac{\mathrm d}{\mathrm dt}
\int\frac{\mathrm d^3k}{(2\pi)^3/V}\int\frac{\mathrm d^3q}{(2\pi)^3/V}
\Big|\int_0^t\mathrm d\tau\langle f|H_W(\tau)|i_{M_F}\rangle\Big|^2=\\
&=\frac{M_\mathrm{GT}^2}{2}\frac{\mathrm d}{\mathrm dt}
\int\frac{\mathrm d^3k}{(2\pi)^3/V}\int\frac{\mathrm d^3q}{(2\pi)^3/V}\nonumber\\
&\hspace{20mm}\times\Big|\int\frac{\mathrm d^3\kappa}{V(4\pi^3\sigma_\kappa^2)^{3/4}}
\,\mathrm e^{\textstyle\,-\frac{\vec\kappa^2}{2\sigma_\kappa^2}}\int\mathrm d^3R
\mathrm\;\mathrm e^{\mathrm i(\vec\kappa-\vec q-\vec k)\vec R}
\sum_jU_{ej}\int_0^t\mathrm d\tau\exp\{\mathrm i\Delta E_jt\}
\Big|^2=\nonumber\\
&=\frac{M_\mathrm{GT}^2}{2}\frac{\mathrm d}{\mathrm dt}
\int\frac{\mathrm d^3k}{(2\pi)^3}\int\frac{\mathrm d^3q}{(2\pi)^3}\nonumber\\
&\hspace{20mm}\times\Big|\int\frac{\mathrm d^3\kappa}{(4\pi^3\sigma_\kappa^2)^{3/4}}
\,\mathrm e^{\textstyle\,-\frac{\vec\kappa^2}{2\sigma_\kappa^2}}
\prod_{i=1}^3\frac{\sin(\frac{\kappa_i-q_i-k_i}{2}L)}{\frac{\kappa_i-q_i-k_i}{2}}
\sum_jU_{ej}\int_0^t\mathrm d\tau\exp\{\mathrm i\Delta E_jt\}
\Big|^2=\nonumber\\
&=\frac{M_\mathrm{GT}^2}{2}\frac{\mathrm d}{\mathrm dt}
\int\frac{\mathrm d^3k}{(2\pi)^3}
\int\frac{\mathrm d^3\kappa}{(4\pi^3\sigma_\kappa^2)^{3/2}}
\int\mathrm d^3\kappa^\prime
\,\mathrm e^{\textstyle\,-\frac{\vec\kappa^2+\vec\kappa^{\prime 2}}{2\sigma_\kappa^2}}
\prod_{i=1}^3\Big[\int\frac{\mathrm dq_i}{2\pi}
\frac{\sin(\frac{\kappa_i-q_i-k_i}{2}L)}{\frac{\kappa_i-q_i-k_i}{2}}
\frac{\sin(\frac{\kappa_i^\prime-q_i-k_i}{2}L)}{\frac{\kappa_i^\prime-q_i-k_i}{2}}\Big]
\nonumber\\&\hspace{20mm}\times
\Big|\sum_jU_{ej}\int_0^t\mathrm d\tau\exp\{\mathrm i\Delta E_jt\}
\Big|^2=\nonumber\\
&=\frac{M_\mathrm{GT}^2}{2}\frac{\mathrm d}{\mathrm dt}
\int\frac{\mathrm d^3k}{(2\pi)^3}
\int\frac{\mathrm d^3\kappa}{(4\pi^3\sigma_\kappa^2)^{3/2}}
\int\mathrm d^3\kappa^\prime
\,\mathrm e^{\textstyle\,-\frac{\vec\kappa^2+\vec\kappa^{\prime 2}}{2\sigma_\kappa^2}}
\prod_{i=1}^3\Big[
\frac{\sin(\frac{\kappa_i-\kappa_i^\prime}{2}L)}{\frac{\kappa_i-\kappa_i^\prime}{2}}
\Big]\Big|\sum_jU_{ej}\int_0^t\mathrm d\tau\exp\{\mathrm i\Delta E_jt\}
\Big|^2=\nonumber\\
&=\frac{M_\mathrm{GT}^2}{2}\frac{\mathrm d}{\mathrm dt}
\int\frac{\mathrm d^3k}{(2\pi)^3}
\int\frac{\mathrm d^3\kappa}{(4\pi^3\sigma_\kappa^2)^{3/2}}
\prod_{i=1}^3\Big[\int\mathrm d\kappa_i^\prime
\frac{\sin(\frac{\kappa_i-\kappa_i^\prime}{2}L)}{\frac{\kappa_i-\kappa_i^\prime}{2}}
\,\mathrm e^{\textstyle\,-\frac{\kappa_i^2+\kappa_i^{\prime 2}}{2\sigma_\kappa^2}}
\Big]\Big|\sum_jU_{ej}\int_0^t\mathrm d\tau\exp\{\mathrm i\Delta E_jt\}
\Big|^2\nonumber
\end{align}

\newpage
\item Detailed derivation of Eq.~(\ref{timDer})\\
\begin{align}\label{TimeIntCalc} 
&\frac{\mathrm d}{\mathrm dt}
\Big|\sum_jU_{ej}\frac{\sin\frac{\Delta E_jt}{2}}{\frac{\Delta E_j}{2}}\,
\mathrm e^{\mathrm i\,\frac{\Delta E_j}{2} t}\Big|^2
=\frac{\mathrm d}{\mathrm dt}
\sum_{jl}U_{ej}U_{el}^*\frac{\sin\frac{\Delta E_jt}{2}}{\frac{\Delta E_j}{2}}\,
\frac{\sin\frac{\Delta E_lt}{2}}{\frac{\Delta E_l}{2}}\,
\mathrm e^{\mathrm i\,\frac{\Delta E_j-\Delta E_l}{2} t}=\nonumber\\
&=\frac{\mathrm d}{\mathrm dt}\Big\{\frac{\mathrm d}{\mathrm dt}\sum_jU_{ej}^2
\frac{\sin^2\frac{\Delta E_jt}{2}}{\big[\frac{\Delta E_j}{2}\big]^2}\,
+\sum_{j\ne l}U_{ej}U_{el}
\frac{\sin\frac{\Delta E_jt}{2}}{\frac{\Delta E_j}{2}}
\frac{\sin\frac{\Delta E_lt}{2}}{\frac{\Delta E_l}{2}}
\mathrm e^{\mathrm i\,\frac{\Delta E_j-\Delta E_l}{2} t}\Big\}=\nonumber\\
&=\frac{\mathrm d}{\mathrm dt}\Big\{2\pi t\sum_jU_{ej}^2\,\delta(\Delta E_j)
+\pi\sum_{j\ne l}U_{ej}U_{el}
\Big[\delta(\Delta E_j)\frac{\sin\frac{\Delta E_lt}{2}}{\frac{\Delta E_l}{2}}+
\frac{\sin\frac{\Delta E_jt}{2}}{\frac{\Delta E_j}{2}}
\delta(\Delta E_l)\Big]\cos\frac{(\Delta E_j-\Delta E_l)t}{2}\Big\}=\nonumber\\
&=\frac{\mathrm d}{\mathrm dt}\Big\{2\pi t\sum_jU_{ej}^2\,\delta(\Delta E_j)
+\pi\sum_{j\ne l}U_{ej}U_{el}
\Big[\delta(\Delta E_j)\frac{\sin\frac{\Delta E_lt}{2}
\cos\frac{\Delta E_lt}{2}}{\frac{\Delta E_l}{2}}+
\frac{\sin\frac{\Delta E_jt}{2}\cos\frac{\Delta E_jt}{2}}{\frac{\Delta E_j}{2}}
\delta(\Delta E_l)\Big]\Big\}=\nonumber\\
&=\frac{\mathrm d}{\mathrm dt}\Big\{2\pi t\sum_jU_{ej}^2\,\delta(\Delta E_j)
+2\pi\sum_{j\ne l}U_{ej}U_{el}\frac{\sin(\Delta E_lt)}{\Delta E_l}
\delta(\Delta E_j)\Big\}=\nonumber\\
&=2\pi\sum_jU_{ej}^2\,\delta(\Delta E_j)
+2\pi\sum_{j\ne l}U_{ej}U_{el}\cos(\Delta E_lt)\delta(\Delta E_j)=\nonumber\\
&=2\pi\sum_j\delta(\Delta E_j)\Big[U_{ej}^2
+\sum_{l(\ne j)}U_{ej}U_{el}\cos(\Delta E_lt)\Big]\nonumber\\.
\end{align}

\item Reduction of the diagonal terms of Eq.~(\ref{decaRateCalc1}) to Eq.~(\ref{decaRateAveFin})\\
\begin{eqnarray}\label{decaRateCalcDiagDetail}
\begin{aligned}
&\lambda_\mathrm{EC}(t)=\\
&=\frac{M_\mathrm{GT}^2}{2}\frac{\mathrm d}{\mathrm dt}
\int\frac{\mathrm d^3k}{(2\pi)^3}
\int\frac{\mathrm d^3\kappa}{(4\pi^3\sigma_\kappa^2)^{3/2}}
\prod_{i=1}^3\Big[\int\mathrm d\kappa_i^\prime
\frac{\sin(\frac{\kappa_i-\kappa_i^\prime}{2}L)}{\frac{\kappa_i-\kappa_i^\prime}{2}}
\,\mathrm e^{\textstyle\,-\frac{\kappa_i^2+\kappa_i^{\prime 2}}{2\sigma_\kappa^2}}
\Big]\sum_jU_{ej}^2\Big|\int_0^t\mathrm d\tau\exp\{\mathrm i\Delta E_jt\}
\Big|^2=\nonumber\\
&=\frac{M_\mathrm{GT}^2}{2}\frac{\mathrm d}{\mathrm dt}
\int\frac{\mathrm d^3k}{(2\pi)^3}
\int\frac{\mathrm d^3\kappa}{(4\pi^3\sigma_\kappa^2)^{3/2}}
\prod_{i=1}^3\Big[\int\mathrm d\kappa_i^\prime\,2\pi
\delta(\kappa_i-\kappa_i^\prime)
\,\mathrm e^{\textstyle\,-\frac{\kappa_i^2+\kappa_i^{\prime 2}}{2\sigma_\kappa^2}}
\Big]\sum_jU_{ej}^2\Big|\int_0^t\mathrm d\tau\exp\{\mathrm i\Delta E_jt\}
\Big|^2=\nonumber\\
&=\frac{M_\mathrm{GT}^2}{2}\frac{\mathrm d}{\mathrm dt}
\int\frac{\mathrm d^3k}{(2\pi)^3}
\int\frac{\mathrm d^3\kappa}{(\pi\sigma_\kappa^2)^{3/2}}\,
\,\mathrm e^{\textstyle\,-\frac{\vec\kappa^2}{\sigma_\kappa^2}}\sum_jU_{ej}^2
\Big|\int_0^t\mathrm d\tau\exp\{\mathrm i\Delta E_j(k,\kappa,\varphi)t\}
\Big|^2=\nonumber\\
&=\frac{M_\mathrm{GT}^2}{2}\int\frac{\mathrm d^3k}{(2\pi)^3}
\int\frac{\mathrm d^3\kappa}{(\pi\sigma_\kappa^2)^{3/2}}\,
\,\mathrm e^{\textstyle\,-\frac{\vec\kappa^2}{\sigma_\kappa^2}}\frac{\mathrm d}{\mathrm dt}
\sum_jU_{ej}^2\Big|\frac{\sin\frac{\Delta E_jt}{2}}{\frac{\Delta E_j}{2}}\,
\mathrm e^{\mathrm i\,\frac{\Delta E_j}{2} t}\Big|^2=\nonumber\\
&=\frac{M_\mathrm{GT}^2}{2}\int\frac{\mathrm d^3k}{(2\pi)^3}
\int\frac{\mathrm d^3\kappa}{(\pi\sigma_\kappa^2)^{3/2}}\,
\,\mathrm e^{\textstyle\,-\frac{\vec\kappa^2}{\sigma_\kappa^2}}
2\pi\sum_j\delta(\Delta E_j)U_{ej}^2
=\frac{M_\mathrm{GT}^2}{2}\int\frac{\mathrm d^3k}{(2\pi)^3}
2\pi\delta(\Delta E)\sum_jU_{ej}^2=\nonumber\\
&=\frac{M_\mathrm{GT}^2}{2}\int\frac{\mathrm d^3k}{(2\pi)^2}\delta(\Delta E)
=\frac{M_\mathrm{GT}^2}{2}\frac{4\pi Q^2}{(2\pi)^2}
=\frac{M_\mathrm{GT}^2Q^2}{2\pi}=\frac{3}{2\pi}[g_AG_FV_{ud}M_\mathrm{EC}Q]^2.
\nonumber
\end{aligned}
\end{eqnarray}

\newpage
\item Reduction of a non--diagonal term of Eq.~(\ref{decaRateCalc1}) to Eq.~(\ref{decaRateOszFin})\\
\begin{eqnarray}\label{decaRateNonDiag}
\begin{aligned}
\lambda_\mathrm{EC}^{jl}(t)&=\frac{M_\mathrm{GT}^2}{2}\int\frac{\mathrm d^3k}{(2\pi)^3}
\prod_{i=1}^3\Big[\frac{\int\mathrm d\kappa_i\int\mathrm d\kappa_i^\prime}
{(4\pi^3\sigma_\kappa^2)^{1/2}}
\,\mathrm e^{\textstyle\,-\frac{\kappa_i^2+\kappa_i^{\prime 2}}{2\sigma_\kappa^2}}
\frac{\sin(\frac{\kappa_i-\kappa_i^\prime}{2}L)}
{\frac{\kappa_i-\kappa_i^\prime}{2}}\Big]\\
&\times U_{ej}U_{el}\frac{\mathrm d}{\mathrm dt}\int_0^t\mathrm d\tau
\exp\{\mathrm i\Delta E_j(k,\kappa^\prime,\varphi^\prime)t\}
\int_0^t\mathrm d\tau
\exp\{-\mathrm i\Delta E_lt(k,\kappa,\varphi)\}=\\
&=\frac{M_\mathrm{GT}^2}{2}\int\frac{\mathrm d^3k}{(2\pi)^3}
\prod_{i=1}^3\Big[\frac{\int\mathrm d\kappa_i\int\mathrm d\kappa_i^\prime}
{(4\pi^3\sigma_\kappa^2)^{1/2}}
\,\mathrm e^{\textstyle\,-\frac{\kappa_i^2+\kappa_i^{\prime 2}}{2\sigma_\kappa^2}}
\frac{\sin(\frac{\kappa_i-\kappa_i^\prime}{2}L)}
{\frac{\kappa_i-\kappa_i^\prime}{2}}\Big]\\
&\times U_{ej}U_{el}2\pi\delta(\Delta E_j(k,\kappa^\prime,\varphi^\prime))
\cos(\Delta E_l(k,\kappa,\varphi)t)=\\
&=\frac{M_\mathrm{GT}^2}{2}\int\frac{\mathrm d^3k}{(2\pi)^3}
\prod_{i=1}^3\Big[\int\frac{\mathrm d\kappa_i}{2\pi}
\,\mathrm e^{\textstyle\,-\frac{\kappa_i^2}{4\sigma_\kappa^2}}
\frac{\sin(\frac{\kappa_i}{2}L)}{\frac{\kappa_i}{2}}\Big]\\
&\times U_{ej}U_{el}2\pi\delta(\Delta E_j(k,0,\varphi))
\cos(\Delta E_l(k,\kappa,\varphi)t)
\end{aligned}
\end{eqnarray}

\item Folding of Gaussians\\
\begin{eqnarray}\label{foldGauss} 
\begin{aligned}
&\int\frac{\mathrm d^3k}{(2\pi\sigma^2)^{3/2}}
\int\frac{\mathrm d^3q}{(2\pi\sigma^2)^{3/2}}\mathrm e^{-\frac{\vec k^2+\vec q^2}{2\sigma^2}}
f(\vec k-\vec q)=
\int\frac{\mathrm d^3k}{(4\pi\sigma^2)^{3/2}}\mathrm e^{-\frac{\vec k^2}{4\sigma^2}}
f(\vec k)
\end{aligned}
\end{eqnarray}

\item Integration of plane waves
\begin{eqnarray}\label{intq} 
\begin{aligned}
&\int\mathrm d^3q\Big|\int\mathrm d^3R\mathrm\;e^{-\mathrm i(\vec q+\vec k)\vec R}\Big|^2=
\int\mathrm d^3q\prod_{i=1}^3\Big|\mathrm\;
\frac{e^{-\mathrm i(q_i+k_i)L/2}-e^{-\mathrm i(q_i+k_i)L/2}}{-\mathrm i(q_i+k_i)}\Big|^2=\\
&=\prod_{i=1}^3\int\mathrm dq_i
\Big[\mathrm\;\frac{\sin(\frac{q_i+k_i}{2}L)}{\frac{q_i+k_i}{2}}\Big]^2
=\prod_{i=1}^3\Big[2\pi L\int\mathrm dq_i\delta(q_i+k_i)\Big]
=(2\pi L)^3\int\mathrm d^3q\,\delta^3(\vec q+\vec k)
\end{aligned}
\end{eqnarray}

\end{itemize}

\end{document}